\documentstyle[prc,aps,manuscript]{revtex}

\begin{document}
 
\title{Influence of the Coulomb Interaction on the Chemical Equilibrium 
of Nuclear Systems at Break-Up\\} 

\author{J. T\~oke, W. Gawlikowicz, and W.U. Schr\"oder\\} 

\address{\it Department of Chemistry, University of Rochester, Rochester, New York
14627\\} 
\maketitle
\bigskip  

\begin{abstract}

The importance of a Coulomb correction to the formalism proposed
by Albergo et al. for
determining the temperatures of nuclear systems at break-up and the 
densities of free nucleon gases is discussed. While the proposed
correction has no effect on the temperatures extracted based on double
isotope ratios, it becomes non-negligible when such temperatures or
densities of free nucleon gases are extracted based on multiplicities
of heavier fragments of different atomic numbers. 

\end{abstract}
 
\bigskip
\bigskip

The formalism by Albergo, Costa, Costanzo, and
Rubbino (ACCR)\cite{albergo} offers simple and elegant
prescriptions for the  experimental determination of the temperature 
$T$, and of the free nucleon densities, $\rho_{nF}$ and  $\rho_{pF}$
in a nuclear system at the instance of the break-up. This formalism
presumes a specific scenario for the decay of excited systems,
similar to the scenarios modeled by the Berlin
Microcanonical Metropolis Monte Carlo\cite{mmmc} and the Copenhagen
Statistical Multifragmentation\cite{smm} models. Therefore, it may be
expected to be meaningful in circumstances (in an excitation energy
domain), where the use of the above two more complete models can be
justified. On the other hand, the ACCR formalism is  incompatible
with models such as the equilibrium-statistical model 
GEMINI\cite{gemini} and the Expanding Emitting Source
Model\cite{eesm}, both of which refer to sequential decay scenarios of
systems of uniform density. The purpose of the present paper is to
point out that in certain circumstances, the approximation of the free
nucleon gas as a collection of non-interacting nucleons, as
assumed in the original ACCR approach, may not be
sufficiently accurate, and that the inclusion of a proper Coulomb 
correction term is warranted. Regardless of the 
magnitude of the effects of such a correction,
its inclusion is warranted already by didactical considerations. 
 
The ACCR formalism\cite{albergo} refers to a fragment production
scenario in which an equilibrated freeze-out/break-up configuration
emerges from initial compression and expansion stages. In this formalism,
the average numbers of fragments or clusters of different mass and
atomic numbers $(A,Z)$ are determined by the requirement of chemical
equilibrium between the fragments and free nucleons - neutrons and
protons (quite obviously, the fragments are
then also in a state of chemical equilibrium among themselves). 
In Ref.~1, the chemical equilibrium between fragments $(A,Z)$
and gases of free neutrons and free protons is described by the equation   
 
\begin{equation}
{\mathaccent "7E \mu_{A,Z}} - B_{A,Z}=Z{\mathaccent "7E \mu_{pF}} + 
(A-Z){\mathaccent "7E \mu_{nF}},
\label{eq:chemeq}
\end{equation}

\noindent
where
$B_{A,Z}$ is the fragment binding energy (taken with positive sign) and
${\mathaccent "7E \mu_{A,Z}}$, ${\mathaccent "7E \mu_{nF}}$, 
and ${\mathaccent "7E \mu_{pF}}$ are the reduced chemical
potentials of a fragment $(A,Z)$, of a free neutron, and a free proton, 
respectively. Here, the qualifier ``reduced'' is used with respect
to the term chemical potential and a tilda is used in the respective
symbolic representation, to distinguish the 
quantities involved in Eq.~\ref{eq:chemeq} from the true
thermodynamical chemical potentials $\mu_{A,Z}$, as defined via the
equation

\begin{equation}
\mu_{A,Z} = [{\partial F(V,T)\over \partial N_{A,Z}}]_{V,T}.
\label{eq:chempot}
\end{equation}

In Eq.~\ref{eq:chempot}, $F(V,T)$ is the free energy of the system at
volume $V$ and temperature $T$. The quantity $N_{A,Z}$ is the average
number of fragments $(A,Z)$, and  the partial derivative
is taken at constant  temperature and volume. Note that, unlike their
``reduced'' counterparts ${\mathaccent "7E \mu_{A,Z}}$, the true
chemical potentials $\mu_{A,Z}$ include not only the binding energy
term $B_{A,Z}$, but also the energy of interaction of the fragments
$(A,Z)$ with the relevant mean Coulomb field. When true chemical
potentials are considered, the chemical equilibrium is expressed
through the following equation:

\begin{equation}
\mu_{A,Z} = Z\mu_{pF} + (A-Z)\mu_{nF} - V_{coul}^{pF}(Z),
\label{eq:chemeq1}
\end{equation}
 
\noindent 
where $V_{coul}^{pF}(Z)$ is the average energy of mutual Coulomb
interaction of $Z$ free protons. The latter Coulomb interaction energy
must be subtracted on the right-hand side of Eq.~\ref{eq:chemeq1} to
compensate for the fact, that in a mean-field type of ``bookkeeping''
of the Coulomb energy that is included in the term $Z\mu_{pF}$, the mutual
Coulomb  interaction energy is double-counted.

For some purposes, one may consider the convenient notion  of a
reduced chemical potential ${\mathaccent "7E \mu (A,Z)}$, which
excludes the fragment  binding energies and the interaction with the
mean Coulomb field. In
Maxwell-Boltzmann statistics, the latter quantity has a simple
relationship\cite{albergo}  to the density $\rho_{A,Z}$ of fragments
$(A,Z)$ representing the number of fragments  per unit break-up
volume:

\begin{equation} 
\rho_{A,Z}={A^{3\over 2}\omega_{A,Z}(T)\over \lambda_T^3}
e^{{\mathaccent "7E \mu_{A,Z}}/T}. 
\label{eq:rhomu} 
\end{equation}

Eqs.~\ref{eq:rhomu} and Eq.~\ref{eq:chemeq} are the two fundamental
equations used in Ref.~[1] to establish the relationship between the
yields of various fragments and the characteristics of a chemically
equilibrated nuclear system at break-up. In Eq.~\ref{eq:rhomu},
$\lambda_{T}=h/\sqrt{2\pi m_oT}$ 
is the nucleon thermal wave-length ($m_o$ is the mass of a nucleon), and
$\omega_{A,Z}(T)$ is the temperature-dependent internal partition
function of fragment $(A,Z)$:

\begin{equation}
\omega_{A,Z}(T)=\Sigma_k(2s_k^{A,Z}+1)e^{-E_k^{A,Z}/T},
\label{eq:partition}
\end{equation}

\noindent
where the summation extends over all bound states
of the fragment $(A,Z)$ with spins $s_k^{A,Z}$ and excitation energies
$E_k^{A,Z}$.

An inspection and comparison of Eq.~\ref{eq:chemeq}, to the more
fundamental Eq.~\ref{eq:chemeq1} reveals a lack of symmetry of the
former equation. While the fragment $(A,Z)$ side of the balance
includes the mutual interaction energy of the constituent  nucleons -
the binding energy $B_{A,Z}$, no equivalent term is present for the
free nucleons on the r.h.s. of 
Eq.~\ref{eq:chemeq}. Yet, the $Z$ free protons do interact among themselves
via long-range Coulomb interactions. Therefore, a more complete
equation for the chemical equilibrium based on 
Eq.~\ref{eq:chemeq1} must include the respective
Coulomb interaction term. Note that, on the fragment side of the balance,
the mutual Coulomb interaction energy of $Z$ protons is included in
the binding energy term $B_{A,Z}$.  One may note also that neither
side of Eq.~\ref{eq:chemeq}  considers  explicitly (or implicitly) the
Coulomb interaction energy of the $Z$ protons with  the remaining
$(Z_{system}-Z)$ ``spectator'' protons of the system.  Such an
omission, however, may be well justified, as these two Coulomb
interaction energies are to a good approximation equal to each other
and, hence, cancel mutually. 

A more complete, Coulomb-corrected equation for the 
chemical equilibrium of fragments $(A,Z)$ and free nucleons, in terms
of reduced chemical potentials ${\mathaccent "7E \mu_{A,Z}}$ 
has the symmetrical form:

\begin{equation}
{\mathaccent "7E \mu_{A,Z}}-B_{A,Z}=
Z{\mathaccent "7E \mu_{pF}}+(A-Z){\mathaccent "7E \mu_{nF}}+
V_{coul}^{pF}(Z).
\label{eq:fulleq}
\end{equation}

\noindent 
Here, $V_{coul}^{pF}(Z)$ represents the average potential
energy of the mutual Coulomb interaction of $Z$ free protons, an
equivalent of the term -$B_{A,Z}$. It is worth noting that, here
(unlike in Eq.~\ref{eq:chemeq1})
the Coulomb interaction term enters with positive sign, as no 
Coulomb interaction is included in the reduced free-proton chemical 
potential ${\mathaccent "7E \mu_{pF}}$.  

While it is clear from simple estimates that the Coulomb correction
term, $V_{coul}^{pF}(Z)$, is of non-negligible magnitude when compared
to typical temperatures of the system, it is not obvious how to
actually evaluate it. A conservative estimate 
for the value of $V_{coul}^{pF}(Z)$ may
be obtained by assuming that this term is equal to the Coulomb
interaction energy of $Z$ protons uniformly distributed over a
spherical volume $V_{free}$ = $Z/\rho_{pF}$, i.e., distributed
with a density equal to that of the gas of 
free protons $\rho_{pF}$ at break-up.
In this case, the correction term is independent of the fragment mass
number $A$:

\begin{equation}
V_{coul}^{pF}(Z)={3\over 5}{e^2Z^2\over ({3Z\over 4\pi\rho_{pF}})^{1/3}}
\approx 1.39Z^{5/3}\rho_{pF}^{1/3} (MeV),
\label{eq:coulcorr1}
\end{equation}

\noindent
where $\rho_{pF}$ is expressed in units of $fm^{-3}$. 

The presence of the Coulomb correction term  $V_{coul}^{pF}(Z)$ in
Eq.~\ref{eq:fulleq} modifies the basic equation 5 of Ref. [1] for the
average number of fragments $(A,Z)$ per unit break-up volume,
$\rho_{A,Z}$. It can now be written more accurately as

\begin{equation}
\rho_{A,Z}={A^{3/2}\lambda_{T}^{3(A-1)}
\omega_{A,Z}(T)\over 2^A}\rho_{pF}^Z\rho_{nF}^{A-Z}
e^{[B_{A,Z}+V_{coul}^{pF}(Z)]/T},
\label{eq:dens}
\end{equation}

\noindent
replacing Eq.~\ref{eq:rhomu}. In Eq.~\ref{eq:dens}, 
$\rho_{nF}$ and $\rho_{pF}$ are the densities (i.e.,
numbers per unit break-up volume) of free neutrons and free protons, 
respectively. A similar result was obtained earlier\cite{shlomo} based on
a more rigorous macrocanonical description of a decaying nuclear 
system in a freeze-out configuration. It is worth noting that, in more 
complete 
theoretical descriptions of equilibrated freeze-out configurations, offered
by the Berlin\cite{mmmc} and Copenhagen\cite{smm} models, the effects
of the Coulomb interaction of the free protons are accounted for in a
rigorous fashion, but remain largely transparent to the model users.

In practical applications of Eq.~\ref{eq:dens}, ratios of properly selected
densities, $\rho_{A_1,Z_1}/\rho_{A_2,Z_2}$, are taken and identified with
the ratios of the respective experimental yields of fragments
$Y_{A_1,Z_1}/Y_{A_2,Z_2}$. Such ratios are free of some model parameters
(e.g., of the densities of free neutron and proton gases, in the
case of double isotope ratios), providing often a simple link between
observable yields and selected characteristics of the break-up
state. 

It is clear from Eq.~\ref{eq:dens} that the introduction of the
Coulomb correction term $V_{coul}^{pF}(Z)$ is of no consequence when
ratios of yields are taken for fragments with identical atomic numbers
$Z$, i.e.,  ratios of experimental fragment yields of the type 
$Y_{A+1,Z}/Y_{A,Z}$. In such cases, the corresponding Coulomb correction
terms for the two isotopes involved cancel each other.  As a result,
this correction has no effect on the outcome of an experimental
evaluation of break-up temperatures based on double-isotope ratios
- the most common use of the ACCR\cite{albergo} approach. A similar
cancellation does not, however, occur in cases when, e.g., an experimental
``thermometer'' is constructed from isotone ratios, 
$Y_{A+1,Z+1}/Y_{A,Z}$, or in cases when relative densities of
free neutron and proton gases, $\rho_{nF}/\rho_{pF}$, are determined
based on an isobaric ratio $Y_{A,Z}/Y_{A,Z+1}$. To assess the significance
of the proposed Coulomb correction, several examples are considered
below. In these examples, it is assumed that $T$=3.3 MeV (as
found\cite{tsang} for the system S+Ag at E/A=22 MeV) and 
$\rho_{pF}=5/(4/3\pi 8.0*141)=0.0011$ fm$^{-3}$ (which corresponds to
5 protons in a break-up volume of radius $R_{break-up}=2.0*141^{1/3}$
fm, as for the system $^{32}$S+$^{109}$Ag). 

First, consider an evaluation of the ratio of
densities of free neutron and free proton gases from the observed
isobaric ratios $Y_{A,Z}/Y_{A,Z+1}$. Based on Eq.~\ref{eq:dens}, one
has

\begin{equation}
{\rho_{nF}\over \rho_{pF}}=R_{raw}F_{Coul}^{n/p},
\label{eq:rhoratio}
\end{equation}

\noindent
where $R_{raw}$ is the value of this ratio deduced
in the absence of the Coulomb
correction (i.e., given by the original ACCR\cite{albergo}
formalism) and $F_{Coul}^{n/p}$ is a correction factor resulting
from the Coulomb term proposed in the present paper:

\begin{equation}
F_{Coul}^{n/p}=e^{[V_{coul}^{pF}(Z+1)-V_{coul}^{pF}(Z)]/T}, 
\label{eq:FCoul}
\end{equation}

Using Eqs.~\ref{eq:FCoul} and \ref{eq:coulcorr1} and the values of 
$T$=3.3 MeV and $\rho_{pF}=0.0011$ fm$^{-3}$, one obtains 
$F_{Coul}^{n/p}$ = 1.1, for the case of the isobaric ratio
$Y_{3,1}/Y_{3,2}$ (tritium - helium-3), $F_{Coul}^{n/p}$ = 1.28, for the
case of the isobaric ratio $Y_{13,6}/Y_{13,7}$, and $F_{Coul}^{n/p}$ =
1.59 when the isobaric ratio $Y_{34,16}/Y_{34,17}$ is utilized. This
example demonstrates that even in the favorable case of light  isobars
$^3H$ and $^3$He, the correction factor is large enough to mandate an
inclusion of the proposed Coulomb correction term in the equation for
the chemical equilibrium. Certainly, this correction factor is quite
sizeable when yields of heavier isobars  are utilized for the
evaluation of the relative densities of free neutron and free proton
gases.

It is worth noting that, according to Eq.~\ref{eq:FCoul}, the Coulomb
correction factor, $F_{Coul}^{n/p}$, 
for the relative densities of gases of free neutrons
and protons is  always greater than unity, since the
exponent $[V_{coul}^{pF}(Z+1)-V_{coul}^{pF}(Z)]$ is
positive. This fact reflects the role of the Coulomb energy in
a neutron-enrichment of the free nucleon gas, a rather trivial
effect that should not be confused with an 
isospin fractionation\cite{serot,pan,chomaz} driven by an
isospin-dependent equation of state of nuclear matter.

As a second example, consider the evaluation of the break-up temperature
$T$ based on a double isotone ratio:

\begin{equation}
R_{isotone}={Y_{A_1,Z_1}/Y_{A_1+1,Z_1+1}\over
Y_{A_2,Z_2}/Y_{A_2+1,Z_2+1}}.
\label{eq:Risotone}
\end{equation}

In such a case, the breakup temperature $T$ is ultimately evaluated 
from the experimentally determined value of the ratio 

\begin{equation}
{\Delta B +\Delta V_{coul}^{pF}\over T} =
{\Delta B\over T_{raw}} = ln(aR_{isotone}),
\label{eq:BtoT}
\end{equation}

\noindent
where $\Delta B =
B_{A_1+1,Z_1+1}-B_{A_1,Z_1}+B_{A_2,Z_2}-B_{A_2+1,Z_2+1}$ and
$\Delta V_{coul}^{pF} = V_{coul}^{pF}(Z_1+1)-V_{coul}^{pF}(Z_1)+
V_{coul}^{pF}(Z_2)-V_{coul}^{pF}(Z_2+1)$. The parameter $a$ in
Eq.~\ref{eq:BtoT} 
accounts for intrinsic
partition functions $\omega$ (see Eq.~\ref{eq:partition})
of the isotones involved,
and $T_{raw}$ is the break-up 
temperature obtained using the
original ACCR\cite{albergo} approach.  

Eq.~\ref{eq:BtoT} allows one to
express the relevant Coulomb correction factor $F_{Coul}^T$ as

\begin{equation}
F_{Coul}^T = {T\over T_{raw}}=
1+{\Delta V_{coul}^{pF}\over \Delta B}.
\label{eq:CoulcorrT}
\end{equation}

Using Eq.~\ref{eq:CoulcorrT}, the value of $\rho_{pF}=.0011$
fm$^{-3}$, and values of binding energies from the mass tables,
one obtains $F_{Coul}^T=0.96$ in the case when the 
experimental double isotone ratio 
$(Y_{13,7}/Y_{12,6})/(Y_{4,2}/Y_{3,1})$ is employed,
$F_{Coul}^T=1.19$ in the case of the isotone ratio
$(Y_{13,7}/Y_{12,6})/(Y_{10,5}/Y_{9,4})$, 
and $F_{Coul}^T=0.31$ in the case of 
the isotone ratio $(Y_{14,8}/Y_{13,7})/(Y_{3,2}/Y_{2,1})$. Again,
the estimated magnitude of the Coulomb correction factor well warrants
an inclusion of the Coulomb correction term in the ACCR formalism.

In summary, the importance of a Coulomb correction term to the
equation for the chemical equilibrium between fragments and the gas of free 
nucleons has been demonstrated. The correction term restores the
symmetry of the equation defining the equilibrium, when the
mutual interaction energies of nucleons in both, bound and
free states, are consistently accounted
for as done with the more rigorous Eq.~\ref{eq:chemeq1}.  
While the proposed correction term has no effect on the
determination of break-up temperatures based on double isotope
ratios and may be small in some cases, its effects on the
determination of break-up temperature from isotone ratios and on the
determination of the relative densities of free proton and neutron
gases may be quite substantial in some other cases. This correction is
certainly important in systematic studies of various 
experimental ``thermometers''
that, by design, include a large variety of  isotonic ratios.     

Useful discussions with S.~Albergo are gratefully acknowledged.

This work was supported by the U.S. Department of Energy grant No.
DE-FG02-88ER40414.

\end{document}